\documentclass[12pt]{iopart}
\usepackage{iopams}
\usepackage{setstack}
\usepackage{graphicx}
\usepackage{html,makeidx}

\begin{document}
\title{Detection Confidence Tests for Burst and Inspiral Candidate Events}
\author{Romain Gouaty for the LIGO Scientific Collaboration}
\address{Louisiana State University, Baton Rouge, LA 70803, USA}
\ead{romain@phys.lsu.edu}

\begin{abstract}
The LIGO Scientific Collaboration (LSC) is developing and running analysis pipelines to search for gravitational-wave transients emitted by astrophysical events such as compact binary mergers or core-collapse supernovae. However, because of the non-Gaussian, non-stationary nature of the noise exhibited by the LIGO detectors, residual false alarms might be found at the end of the pipelines. A critical aspect of the search is then to assess our confidence for gravitational waves and to distinguish them from those false alarms. Both the ``Compact Binary Coalescence'' and the ``Burst'' working groups have been developing a detection checklist for the validation of candidate-events, consisting of a series of tests which aim to corroborate a detection or to eliminate a false alarm. These tests include for example data quality checks, analysis of the candidate appearance, parameter consistency studies, and coherent analysis. In this paper, the general methodology used for candidate validation is presented. The method is illustrated with an example of simulated gravitational-wave signal and a false alarm.
\end{abstract}
\maketitle

\section{Introduction}

The Compact Binary Coalescence (CBC) working group of the LIGO-Virgo joint collaboration is a data-analysis group looking for gravitational-wave signals emitted by inspiralling compact binary systems~\cite{ligo,virgo,thorne,CBCradiation}. The time that such a signal will spend in the frequency bandwidth of the current ground-based gravitational-wave interferometers lies in a range which goes from a few tenths up to several tens of seconds.

The Burst working group of the LIGO-Virgo joint collaboration is a data-analysis group looking for transient gravitational-wave signals without specific assumptions on the expected waveform~\cite{pradier00,zivkovic01}. The typical duration of the signals searched by the Burst group is of the order of the millisecond to a tenth of second.

Both of these working groups use analysis methods that are sensitive to the non-stationary and non-gaussian noise of the detectors. Noise transients can induce false alarm triggers in the analysis pipelines that may result in accidental coincidences between the interferometers. It is therefore crucial to submit each gravitational-wave candidate identified by the analysis to a detection checklist which aims to estimate confidence in this candidate.

The detection checklist is made of a list of standard tests in different stages of development that are used to review the gravitational-wave candidates. An overview of this checklist is provided in Section~\ref{checklistOverview}. Section~\ref{significance} shows a method implemented by the Burst and CBC groups to estimate the statistical significance of the candidates. In Section~\ref{checklistItems}, the paper will describe with more details a few items of the detection checklist, using an example of simulated gravitational-wave signal and false-alarm trigger for illustration purposes.

\section{Overview of the detection checklist \label{checklistOverview}}

A detection checklist to evaluate the significance of candidate-events has been developed by both the CBC and Burst groups. Despite some specificities inherent to the kind of signals that are being looked for by the CBC and Burst searches, the method implemented to estimate confidence in a gravitational-wave candidate is very similar for both data-analysis groups. Thus the summary of the checklist provided in this section applies both to the CBC and Burst groups, unless otherwise noted. 

The list presented below is a short synthesis of the tests implemented in the CBC and Burst detection checklists. As many of these tests are still under development or refinement, the checklist is rapidly evolving, and the following list should not be considered exhaustive. Moreover, it is expected that many of the tests that are at present qualitative and based on the experience gained about the instruments will be developed into quantitative tests. Here we outline the main tests that are currently part of the detection checklist for candidate-events or in the process of implementation:
\begin{itemize}
\item {\bf Statistical significance} The first step of the candidate validation procedure consists of determining the statistical significance of the candidates identified by the analysis pipeline, that is to say the probability of coincident triggers arising from random coincidences of noise triggers. The general method is described in Section~\ref{significance}, which will also explain how the statistical significance of the candidates affects the way the other tests of the checklist are addressed.
\item {\bf Data integrity:} sanity checks to verify that the data set containing the candidate is not corrupted.
\item {\bf Status of the interferometers:} The state of the LIGO interferometers~\cite{ligo} and their sensitivity near the time of the candidate are checked. This test also includes a verification of the data quality flags recorded in the database. Section~\ref{ifoStatus} will show how this test can allow the identification of noisy data segment containing false-alarm triggers.
\item {\bf Environmental or instrumental causes:} We analyze auxiliary channels of the interferometers, such as the environmental sensors or the signals involved in the mirror control loops~\cite{ligo}, to check for the presence of possible noise transients which could be the cause of a false alarm identified by the analysis. This effort is also part of the ``glitch group'' activities~\cite{glitchGroup}. More details on this part of the detection checklist are provided in Section~\ref{instrumentalCause}.
\item {\bf Candidate's appearance:} Part of the detection checklist consists of tests of the candidate's appearance. These tests aim to confirm the presence of a gravitational-wave signal in the data or to identify obvious excesses of noise responsible for a false alarm. A variety of tools are used to examine the data containing the candidate-event, such as time series, time-frequency spectrograms, or the outputs of the search pipeline, namely the signal to noise ratio (SNR) or the $\chi^{2}$~\cite{chisq_bruce_allen} time series in the case of the CBC search~\cite{matchfilter}. The analysis of this graphical information is performed by comparing the results obtained at the time of the candidate-event to the expectations for simulated gravitational-wave signals or for known instances of false-alarm triggers. Two examples of candidate appearance tests will be presented in Section~\ref{appearance}. A complementary and quantitative test to establish a likelihood ranking of the candidates given their estimated parameters is in preparation.
\item {\bf Detection robustness:} The Burst group has developed many independent search algorithms looking for unmodelled gravitational-wave signals, a few instances of which are {\tt Block-Normal}~\cite{blocknormal}, {\tt KleineWelle}~\cite{KleineWelleAndQtransform}, {\tt QOnline}~\cite{qscan} and {\tt Waveburst}~\cite{waveburst}. The robustness of a detection for burst searches is checked by verifying that the candidate-event is identified by different independent algorithms. In the case of the CBC search, the list of candidate-events is obtained from a single analysis pipeline based on match filter algorithm~\cite{matchfilter} using specific inspiral waveforms~\cite{inspiralWaveform}. However a Bayesian analysis dedicated to the search for inspiral signals~\cite{singleMCMC} has also been developed and the CBC group has started using it as an independent algorithm to check the robustness of the detection. Part of the detection robustness test also consists in verifying the accuracy of the detectors' calibration~\cite{calibrationStandard,calibrationHoft} at the time of the detection. A current effort of the CBC group is to automate a test that will check the impact of possible calibration errors on the analysis. To this purpose the data will be reanalyzed using a calibration modified to represent the possible uncertainties on the measurement.
\item {\bf Correlation between interferometers:} Other tests of the detection checklist consist in checking for the correlation between the signals measured in the different detectors of the network composed of the three LIGO interferometers~\cite{ligo}, GEO600~\cite{geo} and Virgo~\cite{virgo}. The Burst group has developed many network or coherent analyses~\cite{externalSearches}, one example of which is {\tt Coherent Waveburst}, based on a constraint likelihood method~\cite{coherentLikelihood,coherentEventDisplay}. The CBC group is developing two tests that check for the expected signal correlations in the data containing coincident triggers from multiple detectors. One of these tests uses the matched filtering algorithm~\cite{matchfilter} to compute the multi-interferometer coherent SNR~\cite{Bose:1999pj,Pai:2000zt} and compares it with the null-stream statistic~\cite{Guersel:1989th} for inspiral waveforms to assess the significance of a trigger being an inspiral event. The essential idea behind this test is that above a certain threshold value for the coherent SNR, real gravitational-wave signals will yield a smaller value for the null-stream statistic than instrumental or environmental glitches of the same coherent SNR. The second test uses a Bayesian approach to infer the posterior distributions of the signal parameters~\cite{coherentMCMC}. Both are currently tested as part of the detection checklist. In addition to these checks for correlation between interferometers, the analysis groups also verify whether the candidate-event is identified by the resonant bar detectors~\cite{aurigaLIGO}.
\item {\bf Coincidence with other types of astrophysical observatories:} The LIGO Scientific Collaboration is implementing procedures to check for possible coincidences between gravitational-wave candidates and Gamma-Ray Bursts, optical transients or neutrinos observations~\cite{externalSearches}.
\end{itemize}
\section{Statistical significance of the candidates\label{significance}}
\begin{figure}[h]
\begin{center}
\includegraphics[width=7.5cm]{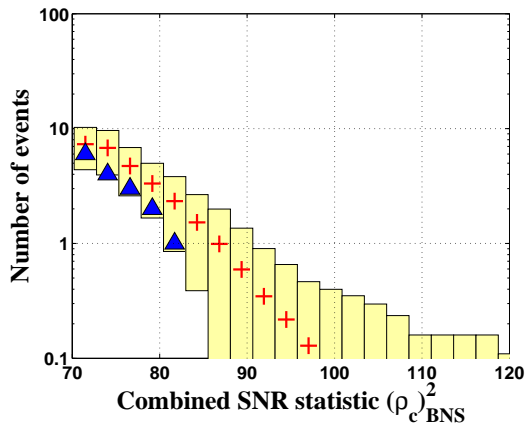}
\caption{\label{S4BNShistogram} Cumulative histogram of the combined SNR, $\rho_{c}$, for the S4 Binary Neutron Star search: for the in-time coincident candidate-events (triangles), and for the estimated background of accidental coincidences (crosses and 1 standard-deviation ranges). This figure has been extracted from \cite{S3S4jointPaper}. All candidates were found consistent with the background.}
\end{center}
\end{figure}
In order to estimate the statistical significance of the candidate-events, the CBC and Burst searches compare the triggers found in coincidence between at least two of the LIGO interferometers to an expected background of accidental coincidences. This background is estimated by repeating the analysis after time-shifting the data of each interferometer with respect to each other. This method, called the time-slides analysis, has already been described in previous publications such as ~\cite{S3S4jointPaper}. Figure~\ref{S4BNShistogram}\footnote{Reprinted figure with permission from Abbott B. {\it et al} (LIGO Scientific Collaboration), {\it Phys. Rev.} D Vol. {\bf 77}, 062002 (2008). Copyright 2008 by the American Physical Society.~\htmladdnormallink{{\bf URL:}~http://link.aps.org/abstract/PRD/v77/e062002}{http://link.aps.org/abstract/PRD/v77/e062002}} shows an example of comparison between the in-time coincident candidate-events and the expected background for the Binary Neutron Star search run over the data taken during the LIGO fourth (S4) science run ~\cite{S3S4jointPaper}. In this example, the background was estimated with one hundred time-slide experiments. The goal of a candidate's follow-up with the checklist depends on whether or not the candidate has a low probability of being an accidental coincidence. If the loudest candidate is consistent with the estimated background (as the candidates shown in figure~\ref{S4BNShistogram}), then the few loudest candidates are investigated to make sure that the accidental coincidences are not due to an undiagnosed instrumental artefact. When such an instrumental artefact is diagnosed properly, this can lower the background and reveal gravitational waves. In case a candidate has a low probability of accidental coincidence (i.e. if the candidate is lying above the estimated background), the goal of the follow-up is then to strengthen our confidence in a possible detection by submitting the candidate to the detection checklist that gravitational-wave signals should pass successfully.

\section{Detailed examples of the checklist\label{checklistItems}}
This section will highlight a few examples of tests used for the review of candidate-events. In order to illustrate the expected results for an inspiral gravitational-wave signal we will refer to a simulated inspiral signal. This simulation was performed by acting on one of the arm test masses of the interferometers to generate a differential motion of the interferometer arm cavities approximately equivalent to the expected effect of a gravitational wave. This simulated gravitational-wave signal was detected by the CBC analysis and stands as an outlier above the estimated background. We will also illustrate the behaviour of the detection checklist when it is applied to an example of false alarm. In the following subsections, we will refer to the simulated gravitational wave as {\it Candidate G} and to the false alarm as {\it Candidate F}.

\subsection{Status of the interferometers} \label{ifoStatus}
\begin{figure}[h]
\begin{center}
\includegraphics[width=0.65\textwidth]{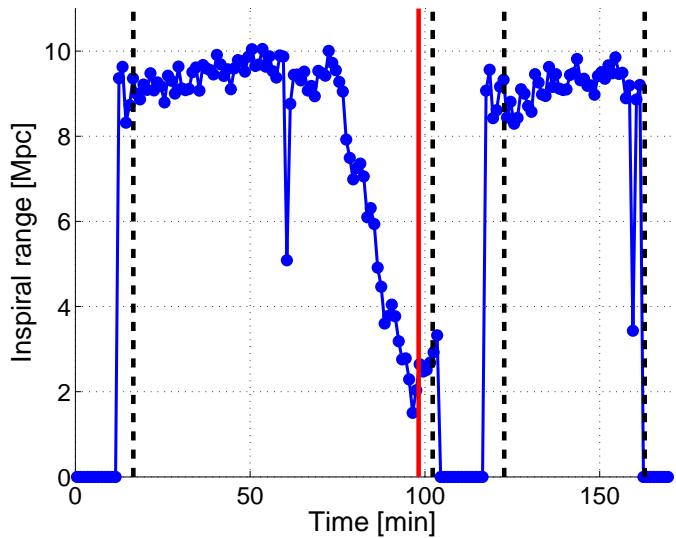}
\caption{\label{BadSenseMon} Inspiral range as a function of time at Livingston (the origin of time is arbitrary). The dashed vertical lines indicate the start and end times of the segments when the interferometer was in science mode. Two science segments are represented in this figure (corresponding to the time windows 16.5-102.2 min and 122.7-162.9 min). The {\it Candidate F} was found in the first science segment, at t$\simeq$98.2~min. This time is underlined by a solid vertical line.}
\end{center}
\end{figure}
A fraction of the detection checklist consists in verifying the status of the interferometers and the data quality in the segment containing the candidate. This includes examining the figures of merit (detectors state, sensitivity, seismic trends) posted in the detectors log, scanning the database which contains the list of data quality flags, as well as checking the information reported by the ``glitch group''~\cite{glitchGroup}. The goal of this study is to check for a possible misbehaviour of the detectors or an unusual excess of noise which could translate into a higher rate of false-alarm triggers and thus reduce our confidence in the candidate-event. For instance we check how the detectors sensitivity varies in time and how it might affect the performances of our searches.

Figure~\ref{BadSenseMon} shows a plot displaying the minute trends of the inspiral search range at the Livingston's site, called {\it inspiral range}~\cite{inspiralRange}, for a time window of about 3 hours which includes the {\it Candidate F}. The {\it inspiral range} is conventionally defined as the distance at which the coalescence of a 1.4-1.4 $M_{\odot}$ binary neutron stars system would be detected by the search with a SNR of 8 averaged over all sky positions and orientations. During the day from which the three hours of data shown in figure~\ref{BadSenseMon} have been extracted, the typical {\it inspiral range} measured in science mode was fluctuating between 8.5 and 10.5~Mpc. However one can notice that the first science segment shown in figure~\ref{BadSenseMon} (between t=16.5~min and t=102.2~min) ends with a sharp drop in the {\it inspiral range} for about twenty five minutes. The inspiral trigger associated with {\it Candidate F} (false alarm) in the Livingston data was found within this segment (it is highlighted by the solid vertical line in figure~\ref{BadSenseMon}) while the {\it inspiral range} was about 2.5~Mpc, that is to say well below the averaged sensitivity reached by the interferometer during this day. At these times, the background is usually significantly larger than the average during the run used to calculate the statistical significance of the candidate. Ranking the science segments by the standard deviation of the inspiral range, this segment was found to be in the top 0.5\%, indicating a very noisy time. Since the {\it Candidate F} was first examined, a data quality flag has been created to mark such noisy segments (this is an advisory flag used to exert caution in case of a detection candidate). The presence of an unusual excess of noise in the Livingston interferometer was also reported in the detector log by the control room experts which were monitoring the detector's behaviour at that time. The remarkable noisiness of the detector prejudices us against {\it Candidate F}, although this check does not prove that the candidate itself is due to detector noise. An evidence of the nature of the candidate is brought by its SNR or {\it $\chi^{2}$}~\cite{chisq_bruce_allen} time series as discussed in Section~\ref{appearance}.
\subsection{Environmental and instrumental causes} \label{instrumentalCause}
\begin{figure}[h]
\begin{center}
\includegraphics[width=0.6\textwidth]{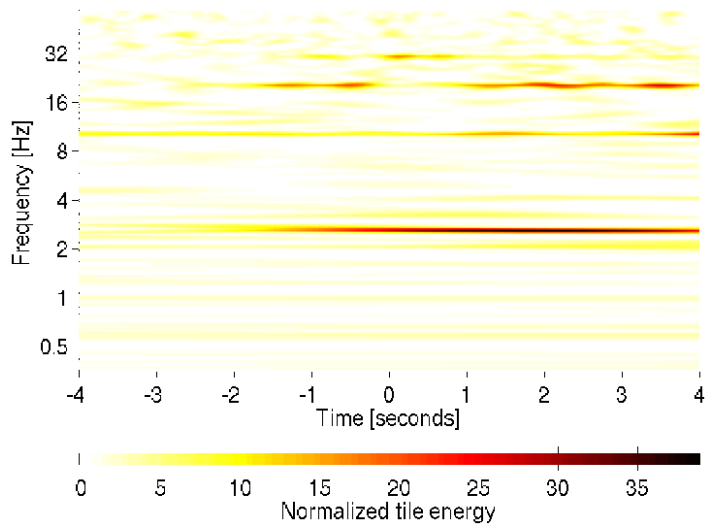}
\caption{\label{seismicQscan} {\it Q spectrogram} of a seismometer channel located near the end mirror test mass of the Hanford 2~km interferometer. The loudest transient found in this spectrogram is located at low frequency (f$\simeq$2.6~Hz) and is time-coincident with the inspiral trigger associated to the {\it Candidate G} (simulated gravitational-wave signal) whose position corresponds to the origin of the x axis (t=0~s).}
\end{center}
\end{figure}
\begin{figure}[h]
\begin{center}
\includegraphics[width=0.6\textwidth]{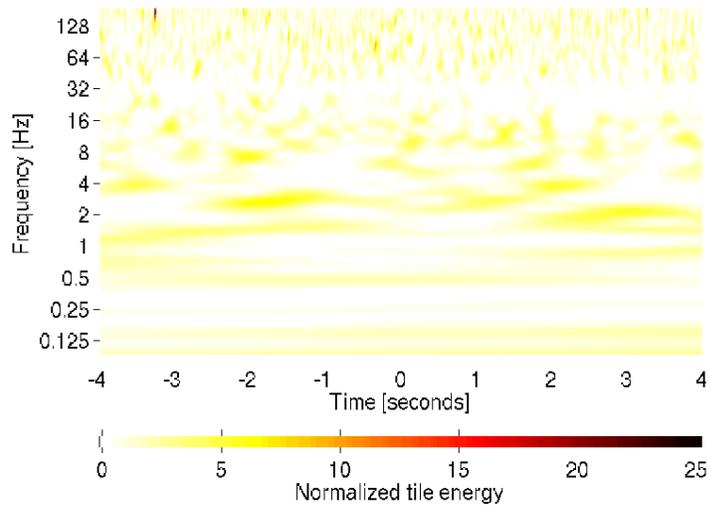}
\caption{\label{DarmQscan} {\it Q spectrogram} of the error signal of the differential mode control loop of the Hanford 2~km interferometer at the time of the detection of {\it Candidate G} (this is the main port sensitive to gravitational waves). The origin of the x axis (t=0~s) is the same as in figure \ref{seismicQscan}.}
\end{center}
\end{figure}
In order to check for possible instrumental artefacts that could be responsible for false alarm triggers, we examine the auxiliary channels of the detectors in a few seconds long window around the candidate-events. For this purpose time-frequency maps of auxiliary channels are being analyzed, using an event visualization tool called {\tt QScan}, which is based on a Q-transform~\cite{qscan}. {\it Qscan} produce time-series and {\it Q spectrograms} of the auxiliary channels in which transients are detected. The {\it Q spectrograms} correspond to time-frequency decompositions using sinusoidal Gaussians characterized by a central time, central frequency, and a quality factor Q. An example of {\it Q spectrogram} is provided in figure~\ref{seismicQscan}, where the examined channel is a seismometer located near the end mirror test mass of the Hanford 2~km interferometer. This kind of channel can measure seismic disturbances at frequencies below a few Hertz. In figure~\ref{seismicQscan} a low frequency transient (approximately at 2.6~Hz) lasting for a few seconds (between t$\simeq$-2~s and t$\simeq$4~s along the time axis) is visible. Moreover this seismometer's transient is coincident with the {\it Candidate G} (simulated gravitational-wave signal) whose time corresponds to t=0~s in figure~\ref{seismicQscan}. When a candidate-event is simultaneous with a transient in an auxiliary channel, further investigations are performed as explained below.\\
\\
\begin{figure}[h]
\begin{center}
\includegraphics[width=0.65\textwidth]{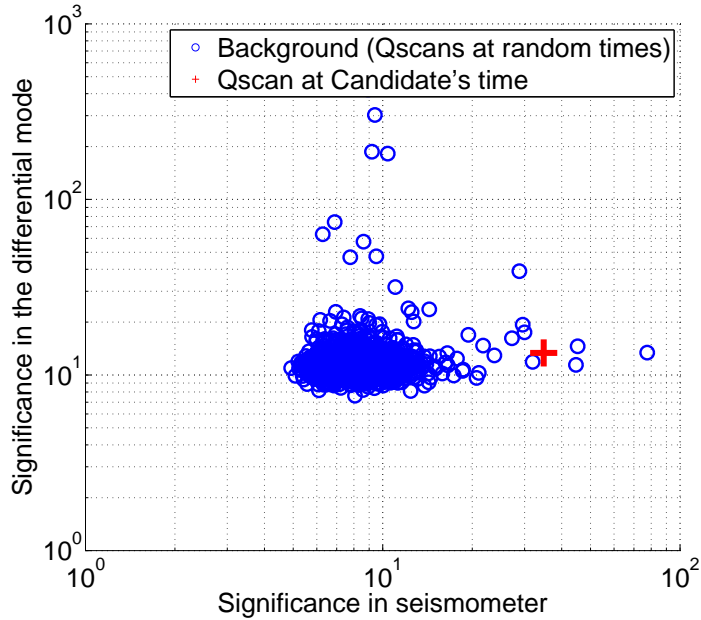}
\caption{\label{scatteredQscan}Scatter plot of {\it QScan significance} in a seismometer channel (x axis) and in the error signal of the differential mode control loop (y axis) of the H2 interferometer. The ``circles'' refer to the {\it QScan significance} measured at times randomly distributed over the first calendar year of the fifth LIGO science run (``S5'')~\cite{ligo} to estimate the seismometer background. The ``cross'' symbol refers to the {\it QScan significance} measured at the time of the {\it Candidate G} (simulated gravitational-wave signal).}
\end{center}
\end{figure}
{\it Statistical significance of instrumental transients}\\
\\
A transient found in an auxiliary channel is characterized by a parameter called {\it Z significance}~\cite{qscan}, equivalent to a SNR squared for the Q-transform. Under the assumption of Gaussian noise, the Z significance could be associated to a false alarm rate proportional to $e^{-Z}$. However, due to the non-Gaussian, non-stationary nature of the noise present in the auxiliary channels, we cannot rely on the previous assumption. Therefore, in order to obtain a statistical ranking of an instrumental transient, we need to determine empirically the background of the auxiliary channel containing this transient.

For each auxiliary channel the background is estimated by running {\tt QScan} at times which are randomly distributed over data-taking epochs and by recording the Z significance measured at each of these random times. This provides a distribution of {\it Z significances} corresponding to the background of the auxiliary channel. Since the background of an auxiliary channel might change during a long data-taking due for example to seasonal variations of the environmental noises or to instrumental drifts, we perform several estimations of the background, over large (such as a year) and short (such as a few days) epochs, and compare the results.

Once the auxiliary channel background has been estimated, we can then compare the Z significance measured at the time of the analyzed candidate to this background, and thus obtain a statistical ranking of the instrumental transient. Figure \ref{scatteredQscan} shows a scatter plot of the {\it Z significance} measured in the seismometer ($x$ axis) versus the {\it Z significance} measured in the error channel of the differential mode control loop of the H2 interferometer ($y$ axis), which is the main port sensitive to gravitational waves. The ``circles'' refer to the background estimated from about 1000 times randomly distributed over the first calendar year of the fifth LIGO science run (``S5'')~\cite{ligo} while the ``cross'' symbol shows the significances measured in both channels at the time when the {\it Candidate G} was detected. A seismic transient whose {\it Z significance} would be comparable with the median of the seismometer background distribution could be ignored as irrelevant. On the contrary the seismic transient shown in figure~\ref{seismicQscan} has a Z significance (read on the $x$ axis) higher than 99.7\% of the background, which makes it statistically relevant. Another estimation of the seismometer background was obtained by analyzing only two days of data including the time of the {\it Candidate G} and led to a similar conclusion.

The fact that a statistically loud seismic transient is found at the time of the {\it Candidate G} is however not sufficient to rule it out as a possible detection. The next question that needs to be addressed is to identify whether or not this environmental transient couples to the interferometer output port.\\
\\
{\it Coupling of environmental disturbances into the interferometer output port}\\
\\
The coupling of an environmental disturbance into the interferometer output port can be proven by comparing the {\it Q spectrogram} of the auxiliary channel to the {\it Q spectrogram} of the interferometer output port and by looking for possible correlations between these two channels.

A high frequency disturbance might couple linearly into the gravitational-wave bandwidth of the output port. Transfer functions have been measured between environmental channels (such as magnetometers, microphones, radio channels) and the interferometer output port~\cite{robertLSC0803}. These transfer functions can be compared to the amplitude ratio measured in the {\it Q spectrograms} of the auxiliary channel and the interferometer output port at the time of the candidate. If the amplitude ratio is consistent with the transfer function, this proves that the environmental disturbance coupled to the output port, and consequently leads to the rejection of the candidate as a possible detection.

In the case of a low frequency seismic transient, noise up-conversion mechanisms might induce a false-alarm trigger in the gravitational-wave bandwidth. One can notice in figure~\ref{scatteredQscan} that the {\it Z significance} measured in the error signal of the interferometer differential mode (read on the $y$ axis) at the time of the {\it Candidate G} is comparable to the significances obtained at random times. Therefore, despite the statistical relevance of the transient in the seismometer, figure~\ref{scatteredQscan} does not argue in favor of a possible coupling into the interferometer output port. Moreover the comparison between the {\it Q spectrograms} of the seismometer (figure \ref{seismicQscan}) and the error signal of the interferometer differential mode (figure \ref{DarmQscan}) does not indicate any correlation between these channels neither at the frequency of the seismic transient (2.6~Hz), neither at higher frequencies. Therefore it is unlikely that the seismic transient be the cause of the inspiral trigger associated to the Candidate G\footnote{Knowing that the {\it Candidate G} is a simulated gravitational-wave signal, the presence of an inspiral trigger is indeed not related to the seismic transient.}.
\subsection{Candidate appearance} \label{appearance}
\begin{figure}[h]
    \begin{minipage}[c]{.5\linewidth}
    \includegraphics[width=7.5cm]{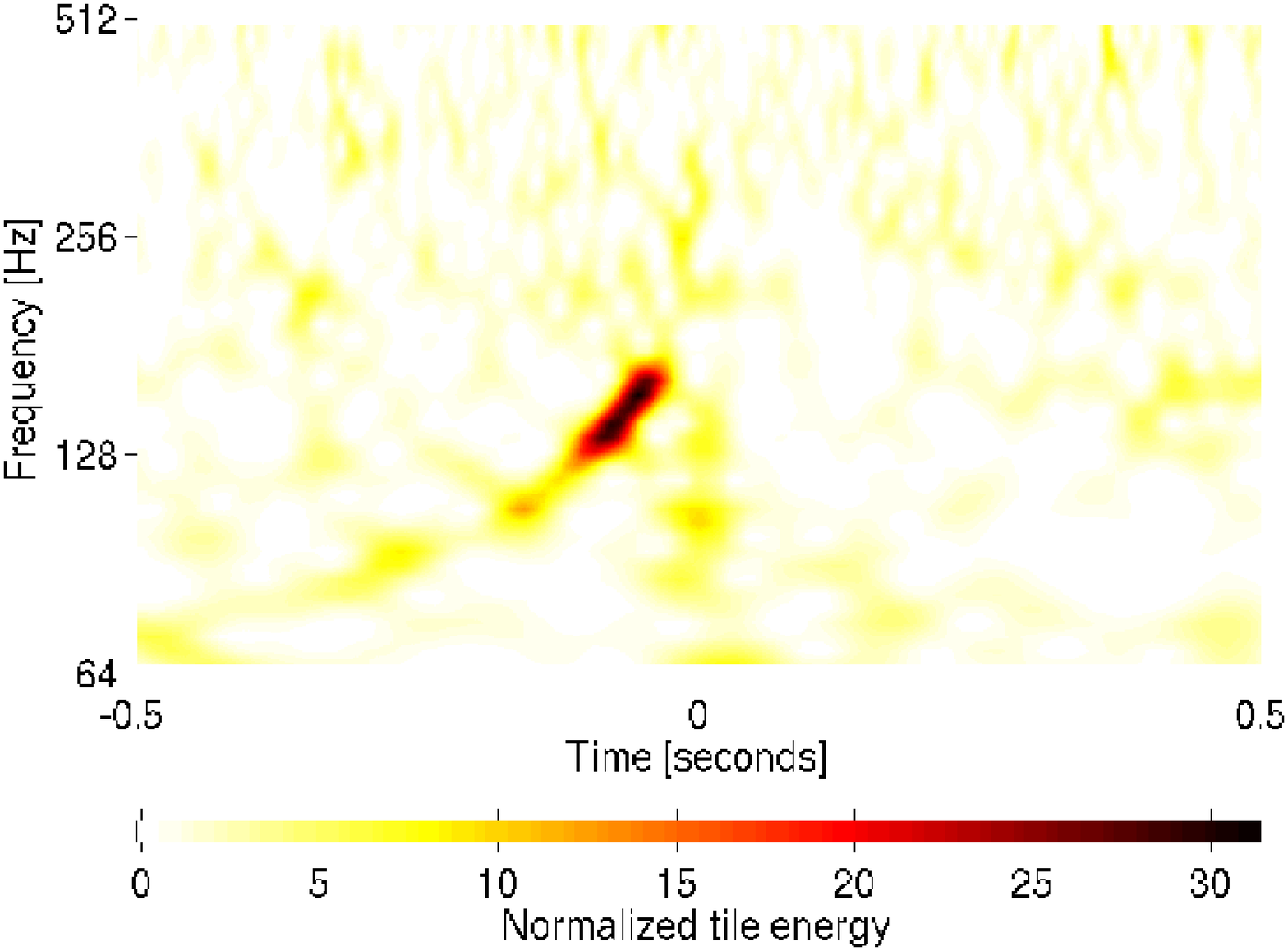}
    \end{minipage} \hfill
    \begin{minipage}[c]{.5\linewidth}
    \includegraphics[width=7.5cm]{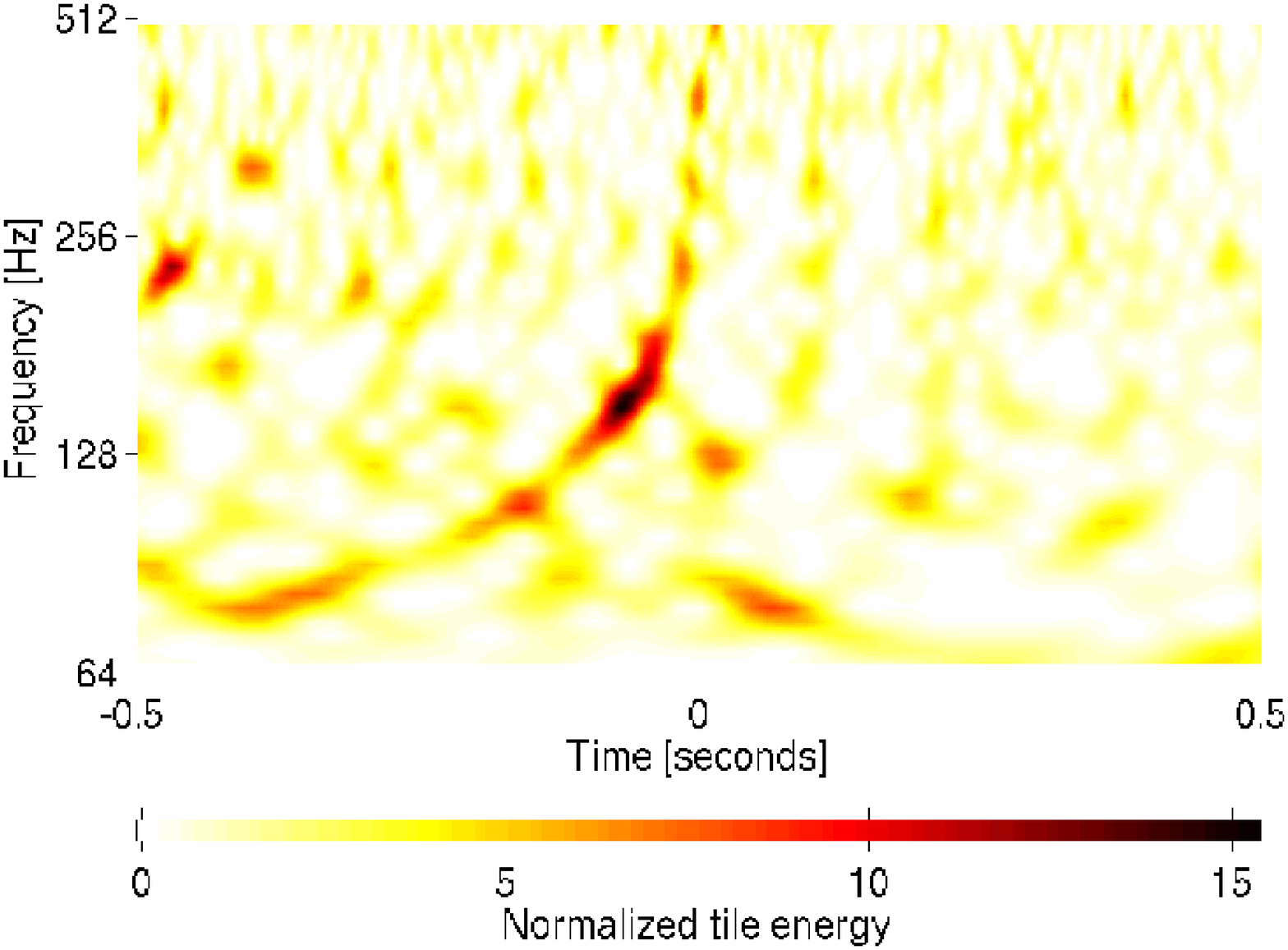}
    \end{minipage}
    \caption{\label{hoftQscan}{\it Q spectrograms} of the data containing the {\it Candidate G} (simulated gravitational-wave signal) in the two more sensitive detectors: Hanford 4~km (left) and Livingston 4~km (right). A ``chirp'' waveform (frequency increasing with time) is visible in the data of the two detectors.}
\end{figure}
In this section two examples of qualitative checks of the candidate's appearance are illustrated: a check of the candidate's time-frequency map, and a check of the output of the match-filtering algorithm~\cite{matchfilter} used to search for inspiral gravitational-wave signals.\\
\\
{\it Q spectrograms of the candidate}\\
\\
A {\tt QScan} of the data in which a candidate-event as been detected is examined in order to perform the following checks:
\begin{itemize}
\item The presence of a possible known signal waveform that might confirm the detection is verified. However, a low SNR inspiral signal is not expected to be visible in a {\it Q spectrogram}. Therefore the absence of visible known waveform in the spectrogram does not rule out a possible detection.
\item The presence of an obvious excess of noise in the data is also checked.
\end{itemize}
Figure~\ref{hoftQscan} shows the {\it Q spectrograms} of the {\it Candidate G} in the two more sensitive interferometers where this simulated gravitational-wave signal was injected. The transient visible in the H1 and L1 data corresponds to the typical ``chirp'' pattern (frequency increasing with time) that is characteristic of an inspiral signal. The simulated gravitational-wave signal is thus visible in these two spectrograms. It is actually not clearly visible in the {\it Q spectrogram} of the H2 data (not represented here) because of the lower SNR in this interferometer.\\
\\
\begin{figure}[h]
    \begin{minipage}[c]{.5\linewidth}
    \includegraphics[width=7cm]{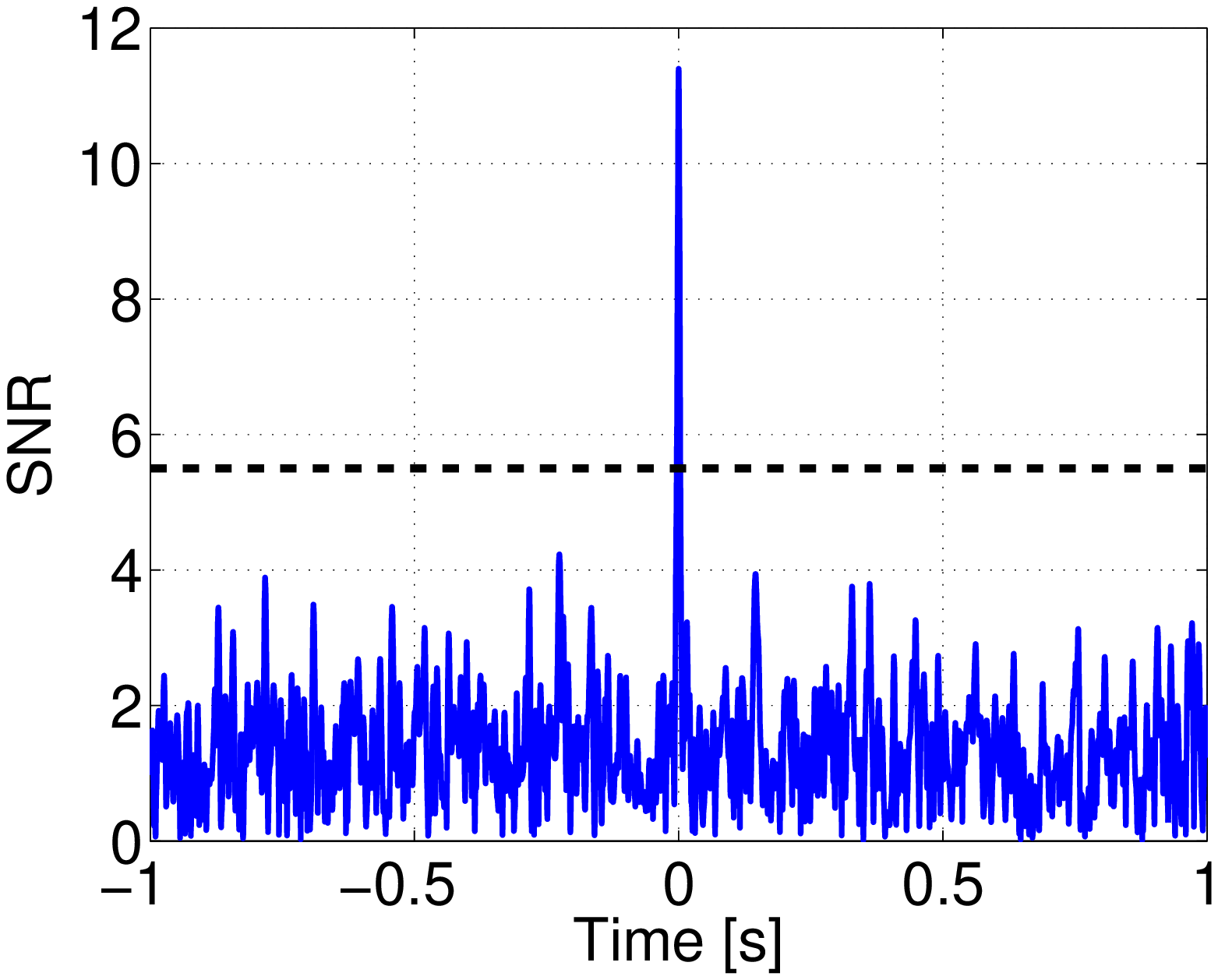}
    \end{minipage} \hfill
    \begin{minipage}[c]{.5\linewidth}
    \includegraphics[width=7cm]{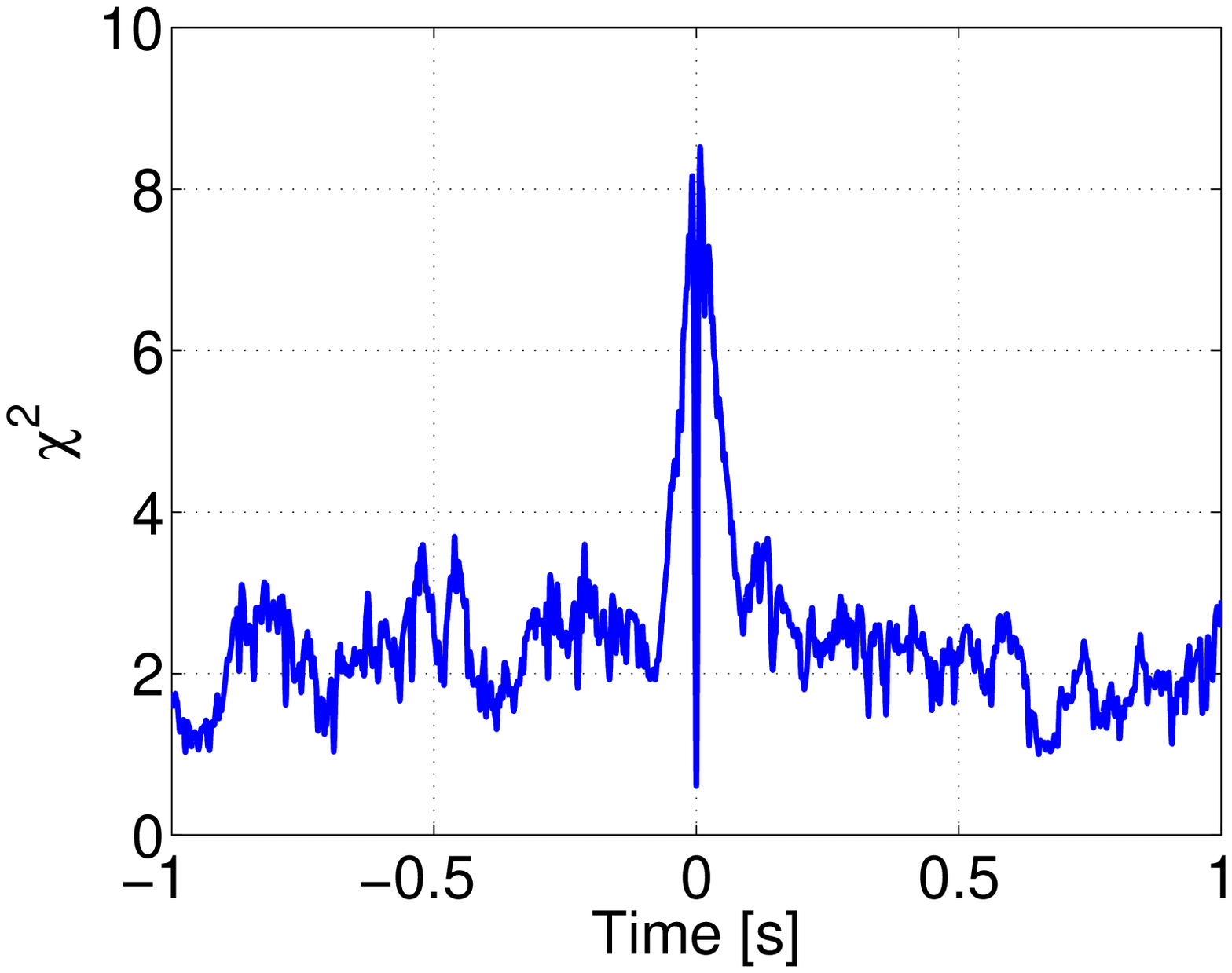}
    \end{minipage}
    \caption{\label{snrCandidateG}SNR (left) and $\chi^{2}$ (right)~\cite{chisq_bruce_allen,matchfilter} time series obtained after match-filtering the Livingston data containing the {\it Candidate G} (simulated inspiral gravitational-wave signal). The time origin on the x axis coincides with the time of the inspiral trigger. In the left plot the dashed horizontal line at SNR=5.5 corresponds to the SNR threshold used for this analysis. SNR peaks exceeding this threshold are recorded as inspiral triggers by the analysis pipeline. In the right plot, the $\chi^{2}$ time series shows a characteristic pattern for a few tens of milliseconds around the time of the {\it Candidate G} (t=0~s) which corresponds to the expectations for a gravitational-wave signal. In particular the $\chi^{2}$ is minimum at t=0~s when the triggered waveform best matches the signal present in the data.}
\end{figure}
\begin{figure}[h]
    \begin{minipage}[c]{.5\linewidth}
    \includegraphics[width=7cm]{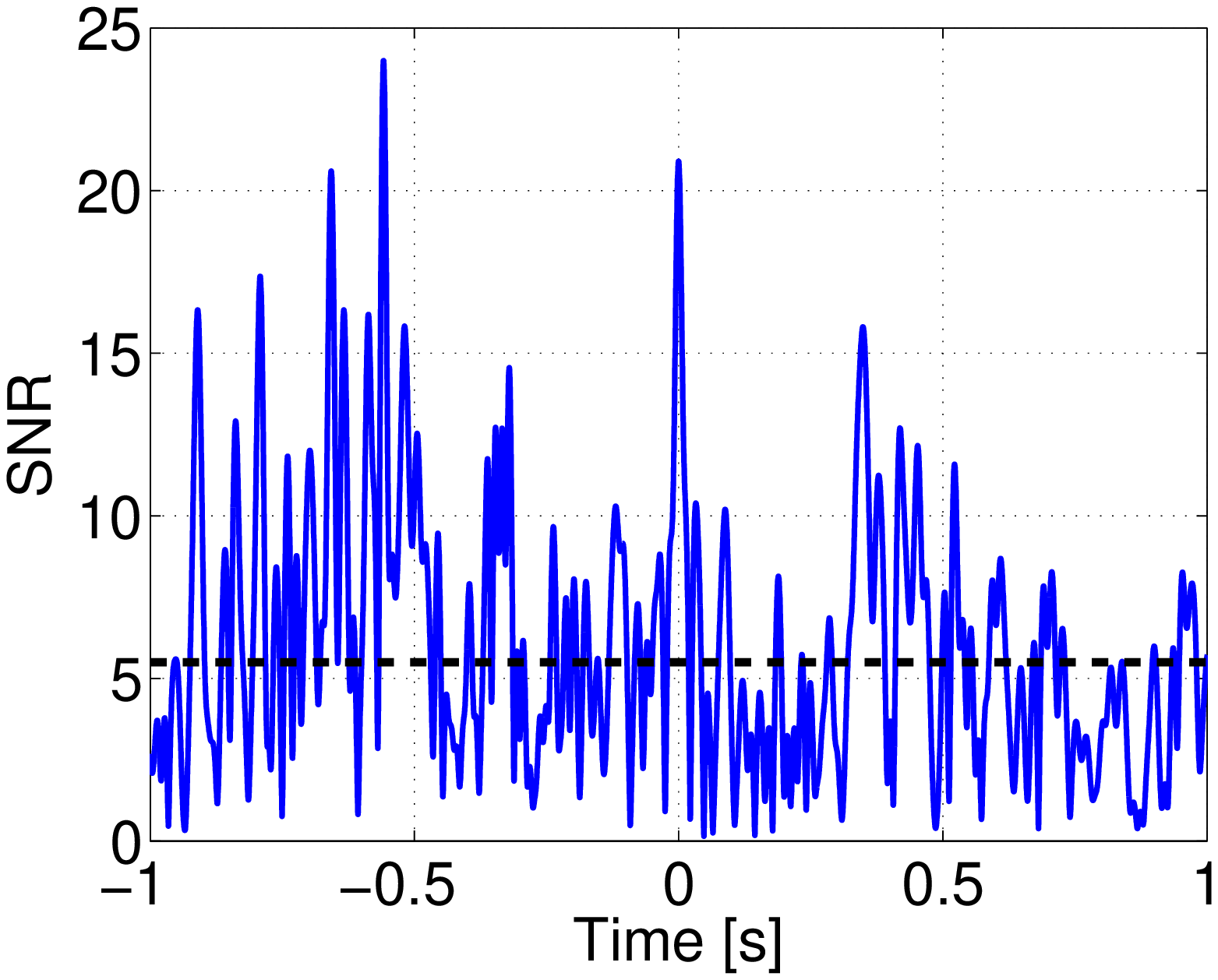}
    \end{minipage} \hfill
    \begin{minipage}[c]{.5\linewidth}
    \includegraphics[width=7cm]{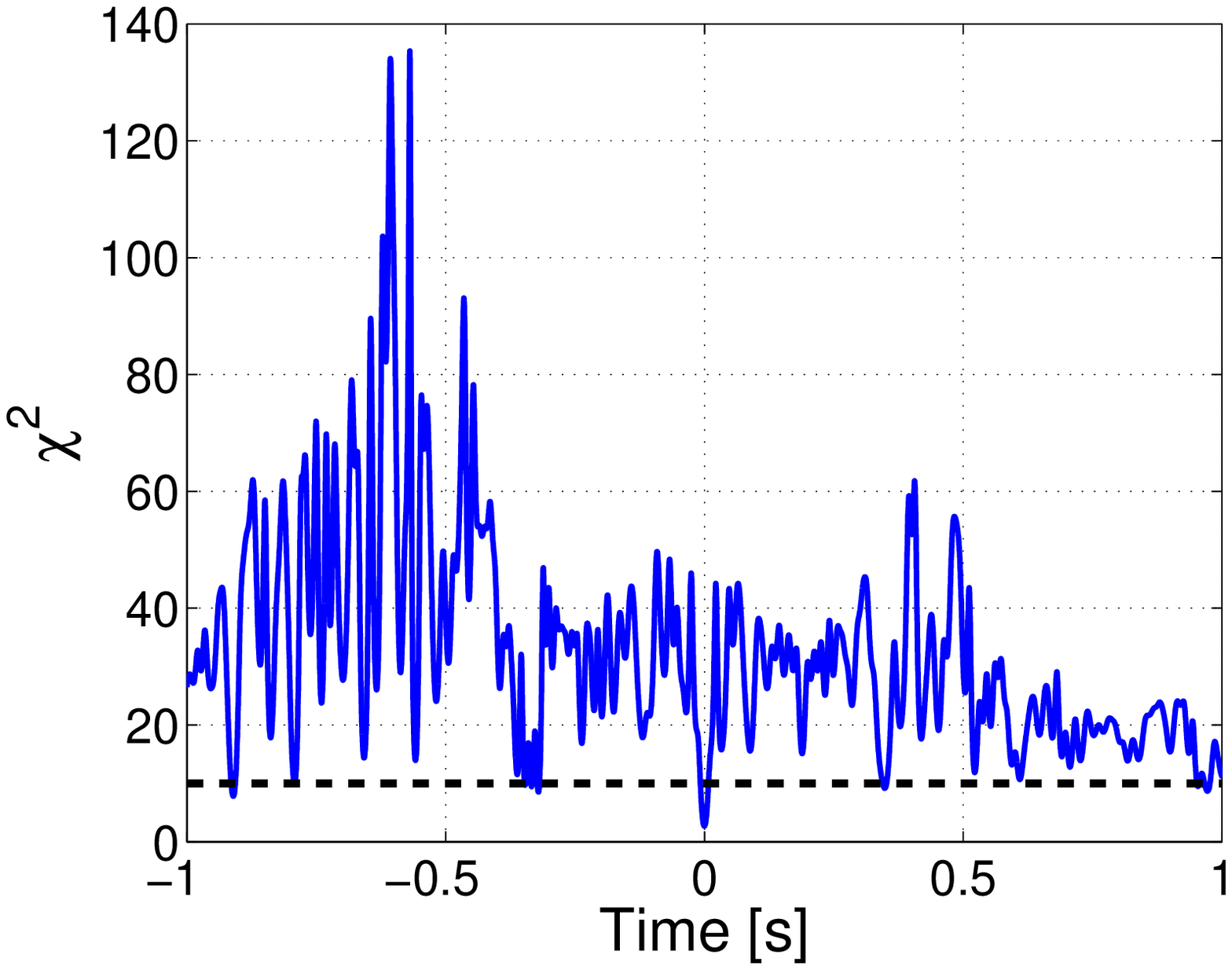}
    \end{minipage}
    \caption{\label{snrCandidateF}SNR (left) and $\chi^{2}$ (right)~\cite{chisq_bruce_allen,matchfilter} time series obtained after match-filtering the Livingston data containing the {\it Candidate F} (false alarm). The time origin on the x axis coincides with the time of the inspiral trigger. In the left plot the dashed horizontal line at SNR=5.5 corresponds to the SNR threshold, while, in the right plot, the dashed horizontal line at $\chi^{2}=10$ corresponds to the $\chi^{2}$ threshold used for this analysis. The inspiral triggers must exceed the SNR threshold while their $\chi^{2}$ must be lower than the corresponding threshold.}
\end{figure}
{\it Output of the match-filtering algorithm}\\
\\
Another example of check for the candidate's appearance that is used by the CBC group consists in examining the time-series of the SNR obtained after match-filtering~\cite{matchfilter} the data with inspiral waveforms~\cite{inspiralWaveform}, as well as the time-series of a $\chi^{2}$ which tests the consistency between the triggered waveform and the signal present in the data.

An example of the expected time-series for a simulated gravitational-wave signal is shown in figure~\ref{snrCandidateG}. On the left plot, the SNR time series shows a short central peak exceeding the threshold at SNR=5.5 used by this search. This SNR peak corresponds to the time of the trigger associated with the simulated inspiral signal. On the right plot, the $\chi^{2}$ time-series presents a very characteristic shape around the time of the inspiral trigger (t=0~s), which corresponds to the expectations for gravitational-wave signal in stationary gaussian noise. A few milliseconds before the time of the inspiral trigger the $\chi^{2}$ value starts increasing, while it falls to a minimum at t=0~s when the triggered waveform best matches the simulated gravitational-wave signal injected in the data. Finally the $\chi^{2}$ time series presents a symmetrical behaviour after t=0~s. 

Notice that a veto based on the value of the $\chi^{2}$ at the time of the inspiral trigger (which would correspond to t=0~s in figure~\ref{snrCandidateG}) is already automatically implemented in the CBC analysis pipeline. This veto rejects any trigger whose $\chi^{2}$ value exceeds a threshold set to $\chi^{2}=10$ in the search that identified {\it Candidate G} and {\it Candidate F}. The tuning of this threshold tends to be very conservative to assure that real gravitational-wave events with waveforms that might differ slightly from the CBC searches templates are not rejected~\cite{tuningPaper}. The qualitative check performed in the detection checklist is complementary to the $\chi^{2}$ veto as it consists in examining the $\chi^{2}$ time series for several tenths of seconds around the inspiral trigger.

Figure~\ref{snrCandidateF} shows the SNR and $\chi^{2}$ time-series around the time of the {\it Candidate F} at Livingston. Multiple peaks of SNR exceeding the threshold are visible in the left plot, which indicates highly non-stationary data. In the right plot the $\chi^{2}$ time series does present a minimum at t=0~s which is the reason why this candidate was not vetoed by the analysis pipeline. However the $\chi^{2}$ time series also shows large values for the whole two seconds window surrounding the candidate, which clearly differs from the plot shown in figure~\ref{snrCandidateG} and indicates a very noisy stretch of data. Accordingly the {\it Candidate F} can be ruled out as a possible detection, which confirms the first suspicions born from the analysis of the {\it inspiral range} in Section~\ref{ifoStatus}.

\section{Conclusions and perspectives\label{conclusion}}
The Burst and CBC groups are pursuing the refinement of their respective detection checklists for candidate-event validation. Part of these tests are still under development and are expected to become more quantitative as experience about the instruments is gained. The groups are currently aiming to automate the detection checklist in order to build a candidate follow-up pipeline which will improve the swiftness of the analysis. The detection checklist is presently being applied to the candidates obtained by the searches analyzing the data taken during the fifth LIGO science run (``S5'')~\cite{ligo}. The detection checklist should play an even more crucial role in the analysis of the future LIGO science runs, for which we expect better detectors' sensitivities and higher probabilities of gravitational-wave detections.

\ack{}
The authors gratefully acknowledge the support of the United States
National Science Foundation for the construction and operation of the
LIGO Laboratory and the Science and Technology Facilities Council of the
United Kingdom, the Max-Planck-Society, and the State of
Niedersachsen/Germany for support of the construction and operation of
the GEO600 detector. The authors also gratefully acknowledge the support
of the research by these agencies and by the Australian Research Council,
the Council of Scientific and Industrial Research of India, the Istituto
Nazionale di Fisica Nucleare of Italy, the Spanish Ministerio de
Educaci\'on y Ciencia, the Conselleria d'Economia, Hisenda i Innovaci\'o of
the Govern de les Illes Balears, the Scottish Funding Council, the
Scottish Universities Physics Alliance, The National Aeronautics and
Space Administration, the Carnegie Trust, the Leverhulme Trust, the David
and Lucile Packard Foundation, the Research Corporation, and the Alfred
P. Sloan Foundation.

This work has also been supported by NFS award PHY0605496 and PHY0355289. This paper was assigned LIGO document number LIGO-P080042-05-Z.


\section*{References}
\begin{thebibliography}{10}
\bibitem{ligo} Abbott B {\it et al} (LIGO Scientific Collaboration) 2007 LIGO: The Laser Interferometer Gravitational-Wave Observatory {\it Preprint} arXiv:0711.3041
\bibitem{virgo} Acernese F {\it et al} (Virgo Collaboration) 2007 {\it Class. Quantum Grav.} {\bf 24} S381-88
\bibitem{thorne} Thorne K S 1987 in {\it 300 years of gravitation} ed S W Hawking and W Israel (Cambridge University Press, Cambridge) p~330.
\bibitem{CBCradiation} Blanchet L, Damour T, Iyer B R, Will C M and Wiseman A G 1995 {\it Phys. Rev. Lett.} {\bf 74} 3515-18
\bibitem{pradier00} Pradier T, Arnaud N, Bizouard M-A, Cavalier F, Davier M and Hello P 2000 {\it Int. J. Mod. Phys.} D {\bf 9} 309-14 ({\it Preprint} arXiv:gr-qc/0001062)
\bibitem{zivkovic01} Zivkovic I and Weinstein A 2001 Tech. Rep. LIGO-T010157-00 LIGO Project {\bf URL:}~http://www.ligo.caltech.edu/docs/T/T010157-00.pdf
\bibitem{glitchGroup} Blackburn L {\it et al}, 2008 The LSC Glitch Group:
Monitoring Noise Transients During the 5th LIGO Science Run {\it Class. Quantum Grav.} in this volume ({\it Preprint} arXiv:0804.0800)
\bibitem{chisq_bruce_allen} Allen B 2005 {\it Phys. Rev.} D {\bf 71} 062001
\bibitem{matchfilter} Allen B, Anderson W G, Brady P R, Brown D A and Creighton J D E 2005 FINDCHIRP: an algorithm for detection of gravitational waves from inspiraling compact binaries {\it Preprint} arXiv:gr-qc/0509116
\bibitem{blocknormal} McNabb J W C, Ashley M, Finn L S, Rotthoff E, Stuver A, Summerscales T, Sutton P, Tibbits M, Thorne K and Zaleski K 2004 {\it Class. Quantum Grav.} {\bf 21} S1705-10 ({\it Preprint} arXiv:gr-qc/0404123)
\bibitem{KleineWelleAndQtransform} Chatterji S, Blackburn L, Martin G and Katsavounidis E 2004 {\it Class. Quantum Grav.} {\bf 21} S1809-18
\bibitem{qscan} Chatterji S 2005 Ph.D thesis, Massachusetts Institute of Technology {\bf URL:}~http://hdl.handle.net/1721.1/34388
\bibitem{waveburst} Klimenko S and Mitselmakher G 2004 {\it Class. Quantum Grav.} {\bf 21} S1819-30
\bibitem{inspiralWaveform} Blanchet L, Iyer B, Will C M and Wiseman A G 1996 {\it Class. Quantum Grav.} {\bf 13} 575
\bibitem{singleMCMC} R\"{o}ver C, Meyer R and Christensen N 2006 {\it Class. Quantum Grav.} {\bf 23} 4895-906
\bibitem{calibrationStandard} Landry M for the LIGO Scientific Collaboration, 2005 {\it Class. Quantum Grav.} {\bf 22} S985-94
\bibitem{calibrationHoft} Siemens X, Allen B, Creighton J, Hewitson M and Landry M 2004 {\it Class. Quantum Grav.} {\bf 21} S1723-36 ({\it Prepint} arXiv:gr-qc/0405070)
\bibitem{geo} L\"{u}ck H {\it et al} 2006 {\it Class. Quantum Grav.} {\bf 23} S71-78
\bibitem{externalSearches} Abbott B {\it et al} (LIGO Scientific Collaboration and Virgo Collaboration) 2008 {\it Class. Quantum Grav.} {\bf 25} 114051
\bibitem{coherentLikelihood} Klimenko S, Mohanty S, Rakhmanov M and Mitselmakher G 2005 {\it Phys. Rev.} D {\bf 72} 122002
\bibitem{coherentEventDisplay} Mercer R A and Klimenko S 2008 Visualising GW
Event Candidates Using the Coherent Event Display, {\it Class. Quantum Grav.}
in this volume
\bibitem{Bose:1999pj} Bose S, Pai A and Dhurandhar S V 2000 {\it Int. J. Mod. Phys.} D {\bf 9} 325-29 ({\it Preprint} arXiv:gr-qc/0002010)
\bibitem{Pai:2000zt} Pai A, Dhurandhar S and Bose S 2001 {\it Phys. Rev. D} {\bf 64} 042004 ({\it Preprint} arXiv:gr-qc/0009078)
\bibitem{Guersel:1989th} Guersel Y, Tinto M, 1989 Phys. Rev. D {\bf 40}, 3884
\bibitem{coherentMCMC} R\"{o}ver C, Meyer R and Christensen N 2007 {\it Phys. Rev.} D {\bf 75} 062004
\bibitem{aurigaLIGO} Baggio L {\it et al} (AURIGA Collaboration and LIGO Scientific Collaboration) 2008 {\it Class. Quantum Grav.} {\bf 25} 095004 ({\it Preprint} arXiv:0710.0497)
\bibitem{S3S4jointPaper} Abbott B {\it et al} (LIGO Scientific Collaboration) 2008 {\it Phys. Rev. D} {\bf 77} 062002, \htmladdnormallink{{\bf URL:}~http://link.aps.org/abstract/PRD/v77/e062002}{http://link.aps.org/abstract/PRD/v77/e062002}
\bibitem{inspiralRange} Finn L S and Chernoff D F 1993 {\it Phys. Rev.} D {\bf 47} 2198
\bibitem{robertLSC0803} Schofield R 2008 Environmental Coupling During S5, LIGO-G080209-00 LIGO Project {\bf URL:}~http://www.ligo.caltech.edu/docs/G/G080209-00/G080209-00.pdf
\bibitem{tuningPaper} The LIGO Scientific Collaboration 2007 Tuning matched filter searches for compact binary coalescence, LIGO-T070109-01 LIGO Project {\bf URL:}~http://www.ligo.caltech.edu/docs/T/T070109-01.pdf
\end{thebibliography}
\end{document}